\documentclass[11pt,twoside]{article}

\usepackage{asp2006}
\usepackage{epsfig}
\usepackage{graphicx}

\begin{document}
\title{Self-consistent models of triaxial cuspy galaxies with dark 
matter halos}   
\author{R. Capuzzo-Dolcetta$^{(1)}$, L.Leccese$^{(1)}$, D. Merritt$^{(2)}$, A. Vicari $^{(1)}$}   
\affil{$^{(1)}$ Dep. of Physics, Universit\'a di Roma «La Sapienza», Roma, Italia; $^{(2)}$ Rochester Inst. of Technology, Rochester, U.S.A.}    

\begin{abstract} 
We have constructed realistic, self-consistent models of triaxial 
elliptical galaxies embedded in triaxial dark matter halos. 
Self-consistent solutions by means of the standard orbital superposition
technique introduced by Schwarzschild were found in each of the three 
cases studied. Chaotic orbits were found to be important 
in all of the models, and their presence was shown to imply a possible slow
evolution of the shapes of the halos. The equilibrium velocity distribution is reproduced by a Lorentzian 
function better than by a Gaussian. Our results demonstrate for the first 
time that triaxial dark matter halos can co-exist with triaxial galaxies. 
\end{abstract}


\section*{Models and Methods}   
We report briefly here some of the results extensively presented and discussed 
in \citet{cdetal07}.
We considered three different mass models for the galaxy+halo system. 
In all the three cases considered we adopted triaxial shapes for both 
components, but with a different radial distribution for the dark and 
luminous matter (LM). In the first model the dark matter (DM) was assumed to have
the same axial ratios as the luminous matter (0.7:0.86:1), while the second and 
third models were characterized by halos that were more prolate (0.5:0.66:1)
and more oblate (0.7:0.93:1), respectively, than the luminous matter. 

In the absence of analytic expressions for the orbital integrals, one 
must resort to numerical methods to construct self-consistent models 
of galaxies or dark-matter halos. At this scope, 
we followed the \citet{schw79} scheme, which is also often referred to as 
{\it orbital~ superposition} technique. 
We generalized Schwarzschild's method to the construction of two-component
(luminous and dark matter), self-consistent models as follows. First, 
we constructed a catalog of orbits integrating the equations of motion 
for a particle moving in the potential generated by the sum of the 
luminous and dark matter density distributions. Then, two different 
grids of cells, one for the luminous and one for the dark matter 
components, were built.

\section*{Results}
\begin{figure} [h] 
\centering
\includegraphics[angle=0,width=11cm]{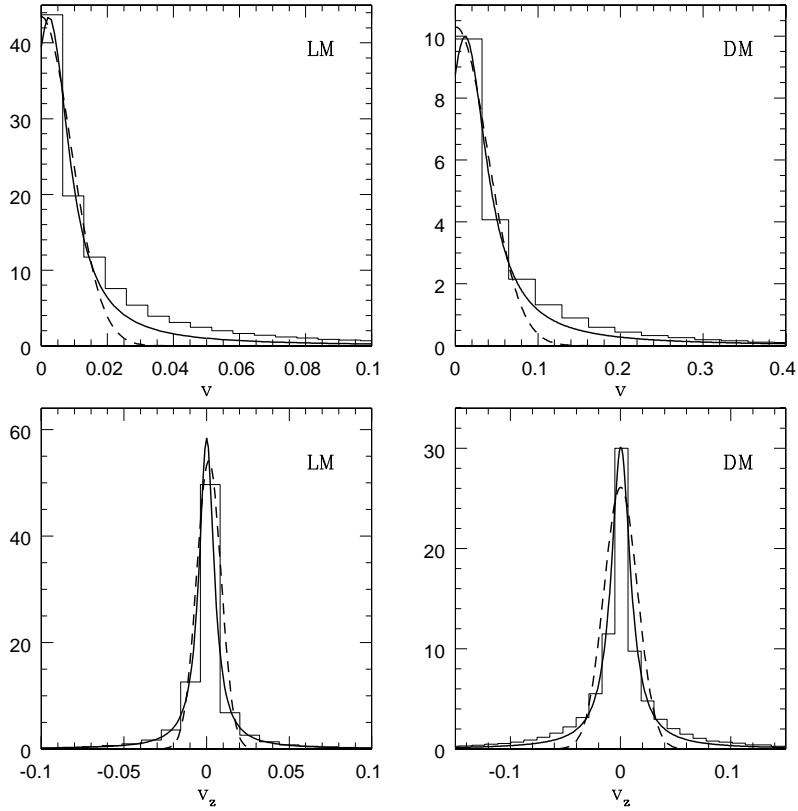}
\caption{Histograms of the velocity distributions of luminous matter (LM, left column) 
and dark matter (DM, right column). Upper panels refer to the distribution of the absolute 
value of the velocity; lower panels refer to the distribution of the z-component of the velocity. 
Solid and dashed curves are the best Lorentzian and Gaussian fits, respectively.} 
\label{fig1} 
\end{figure}

We were able to successfully construct self-consistent models 
in all of the three cases considered for the shape of the DM halo. 
A large fraction of orbits in all the models are chaotic; nevertheless, 
the evolution of the models' shapes over a Hubble time is likely 
to be small.
This result suggests that DM haloes can coexist with triaxial galaxies.
The velocity distribution are non-gaussian in shape; the distribution
is more peaked around the maximum and with broader tails, so to be
better represented with a Lorentzian profile (see Fig. \ref{fig1}). This 
finding suggests care about unquestioned use of Gaussian distributions 
as initial conditions for simulations.

\end{document}